\documentclass[onecolumn,english,reprint,linenumbers]{revtex4}
\usepackage[T1]{fontenc}
\usepackage[latin9]{inputenc}
\usepackage{textcomp}
\usepackage{amsmath}
\usepackage{graphicx}
\usepackage{esint}

\makeatletter

\providecommand{\tabularnewline}{\\}

\@ifundefined{textcolor}{}
{%
 \definecolor{BLACK}{gray}{0}
 \definecolor{WHITE}{gray}{1}
 \definecolor{RED}{rgb}{1,0,0}
 \definecolor{GREEN}{rgb}{0,1,0}
 \definecolor{BLUE}{rgb}{0,0,1}
 \definecolor{CYAN}{cmyk}{1,0,0,0}
 \definecolor{MAGENTA}{cmyk}{0,1,0,0}
 \definecolor{YELLOW}{cmyk}{0,0,1,0}
}

\usepackage{babel}

\makeatother

\usepackage{babel}
\begin{document}

\title{Elastic constants and stress-strain in thin films: application in fiber-textured gold film by X-Ray diffraction }

\author{Edson M. Santos, D. Faurie }

\email{emascarenhassantos@gmail.com;
  faurie@univ- paris13.fr}
%\email{damien@univ paris13.}

\affiliation{Universidade Estadual de Feira de Santana, Av. Transnordestina s/n,
Novo Horizonte, Campus Universitario, Feira de Santana-BA, 44036-900,
Brasil.}

\affiliation{LSPM-CNRS, Universit\'{e} Paris XIII, Sorbonne Paris Cit\'{e}, 93430 Villetaneuse,
France; LFM-DF.}

\date{May, 25$^{th}$ 2018 }

\begin{abstract}

The purpose of the present article is to make a model using analytical equations, based on elasticity theory of continuous media for small deformations,
 with the aim of completely characterizing the material in their mechanical properties as well as the principal stresses-strains of thin films. The approach 
differs from the standard literature which usually brings crystal symmetries or is directly concerned with crystalline materials. It is entirely possible
 to define and to analyze anisotropy in elastic media from first principles in thin films. Therefore, the constitutive relation between strain and
 stress will be considered orthotropic, obeying the generalized Hooke's law. A new equation for the stress of the film-substrate system is proposed
 based on Newton's laws and energy conservation. As an application, it use the technique and data developed
 by Faurie et al (2005) in fiber-textured gold film deposited onto Kapton substrate by combining synchrotron X-Ray diffraction in situ tensile testing.
 As the gold thin film and substrate are considered transversely isotropic, therefore, is firstly required a texture analysis with the purpose of determining
 the possible Euler angles ($\Psi$) and ($\Phi$) for each crystallographic direction that can be used in the model equations. With the data, it is possible to make graphics,
 $\varepsilon$ (strain) X F, for every force applied to the sample. Comparing the experimental graphs with the theoretical equations it was possible obtained their mechanical properties
 and the principal stresses-strains of the anisotropic gold thin film. The results are compared with the results of Faurie et al (2005).

\end{abstract}

%\begin{keywords}
%       keywords - Elasticity moduli, stress, orthotropic material, transversely isotropic material, thin films .
%\end{keywords}

\maketitle

\section{INTRODUCTION}

Given the fast growth and new challenges of nanotechnology, metallic
and non-metallic thin films have a great technological application,
for example, metallic films are typically employed as coatings and
interconnections in microelectronic , solar cells, optical waveguides,
photolithographic masks, solid state devices, among many others\cite{key-1}.
Thus, the studies of the physical properties of thin films have caught 
the attention of the scientific community nowadays.
The study of mechanical properties such as Young's modulus,  shear modulus and Poisson's ratio  are important in materials
science  \cite{key-2} . Measurements of the elastic constants of thin films have
been problematic and usually require sophisticated equipment
or destructive techniques\cite{key-3} being a challenging research topic\cite{key-4} .
A variety of techniques have been used to measure elastic constants
of thin films. Basically, the techniques can be classified into two
groups: destructive and non-destructive. The destructive techniques damage
or modify the sample film properties being tested.The most common techniques
are the tensile tests, technical deflection and resonance ultrasound spectroscopy.
 Non-destructive techniques involve measurements that do
not alter the properties of the films, the substrate
curvature technique with the X-ray diffraction technique. The X-ray diffraction technique
 uses the phenomenon of diffraction in the incidence
of a wave upon the sample by means of an elastic scattering. It is
a high precision technique, being one of the best choices and tools
for measuring the elastic constants of thin films\cite{key-5}. However, the
measurement of the elastic constants in thin films by XRD cannot be
determined directly. It is necessary to measure some property of the
material (such as deformation)\cite{key-6}. We will make use of the
technique and data developed by Faure et al (2005) in gold thin films deposited
onto Kapton substrate, which was studied by using in situ tensile tester
in a four-circle goniometer, on a synchrotron beam line(LURE, France)\cite{key-7},
as well as the model proposed by Santos in his Phd dissertation \cite{key-8}. A new
equation for the stress of the film-substrate system will be deduced. This
technique consists of  measuring the deformation suffered by the material
under uniaxial tension to be applied "in situ".  However, first of all,
it is important to begin with an anisotropy analysis through the pole
figures (measure of the intensity of diffraction planes on the basis
of the angles Euler ($\Psi$) and ($\Phi$) with the purpose of determining 
the possible ($\varPsi$) angles for each crystallographic direction
that can be used in the model equations. Once the angles of the 
crystallographic directions are chosen, it is possible to find out the deformation
for each force applied to the sample.
The determination of the elastic constants as well as the
stress-strain in thin films are obtained by using the elasticity theory of continuous media for small deformations. The constitutive
relation between strain and stress is considered linear, using the
generalized Hooke's law with orthotropic anisotropy.The equation for
the stresses of film-substrate system is proposed based on Newton's
laws and energy conservation. Our results are compared with the
results of Faurie et al (2005). This work is organized as follows:
In \S  II we review linear elasticity theory and the constitutive relation
between stress and strain for orthotropic materials. In \S  III we present our proposal. In \S  IV
we give an application in fiber-textured gold film. In \S  V we present our results and discussion
and in \S  VI  we present our conclusions.

\section{THEORETICAL BACKGROUND}

\subsection{GENERAL REMAKS.}

In the development of a macroscopic phenomenological theory, the use of constitutive relations has a key role. They allow us to characterize the state of macroscopic physical systems experimentally.
A constitutive relation is a relationship between two physical quantities that is specific to a materia. Constitutive relations are particular to each material and are used to classify different materials according to their behavior. The mechanical behavior,
relate tension with a parameter of the body's movement, usually the deformation or deformation rate. There are many other types of constitutive equations such as (I) those that relate stress with deformation and temperature, (II) those that relate stress with electric or magnetic fields, (III) Ohm's law, (IV) the law governig the friction force, etc \cite{key-8}. 

According to Helbig (1994) \textquotedbl{}Historically the study of anisotropic
elastic materials has been synonymous with the study of crystals "  \cite{key-9},
and according to L. Bos (2004)  \textquotedbl{}This historical association does not stem from any sort of conceptual
necessity. On the contrary, it is entirely possible to define and
to analyze anisotropy in elastic media from first principles without any mention of crystals or of crystal symmetry. Pedagogically
this makes it clear that the notion of anisotropy rests solely on the
equations of elasticity, and does not involve any additional assumptions
to the effect that an elastic medium can be modeled as a piecewise collection of crystals, which it cannot. This approach differs from the standard literature
which typically invokes crystal symmetries as a starting point for the analysis of anisotropy, or it is directly concerned with crystalline
materials". Thus, we will begin the studies using the constitutive relation given by the generalized Hooke's law.  For any given orthogonal transformations of three-dimensional space the type of symmetry are triclinic, monoclinic, orthotropic,
 tetragonal, transverse isotropy and isotropy \cite{key-10}. In this work we assume
orthotropic material.

\subsection{LINEAR ELASTICITY THEORY.}

\paragraph{\textup{Linear elasticity is a theory used to study the behavior
of material bodies that are deformed when submitted to external actions
(forces caused by the contact with other bodies, gravitational force
acting on mass, etc.) then returned to their original form when the
external action is removed, with no permanent change in the material
volume. Within certain limits that depend on the material and temperature,
the applied stress is roughly proportional to the strain. In general,
the deformation of a solid involves the combination of a volume change
and a change in the sample\textquoteright s form. Thus, for a given
state of deformation, it is necessary to determine the contributions
of changes in volume and shape.The volume change is caused by variation
of the principal stresses and shape change of the sample is caused
by tangential stresses. In the model, small deformation of the tangential
stresses are neglected and the principal stresses are considered, represented
on laboratory coordinate system by \cite{key-11}:}}

\begin{equation}
\varepsilon_{\varPhi\varPsi}=\varepsilon_{xx}\cos^{2}\text{\ensuremath{\Phi}}\sin{}^{2}\psi+\varepsilon_{yy}\sin^{2}\text{\ensuremath{\Phi}}\sin^{2}\psi+\varepsilon_{zz}\cos^{2}\psi
\end{equation}
Where $ \Phi $ and $ \Psi $ are Euler angles.

\begin{figure}
\includegraphics[scale=0.5]{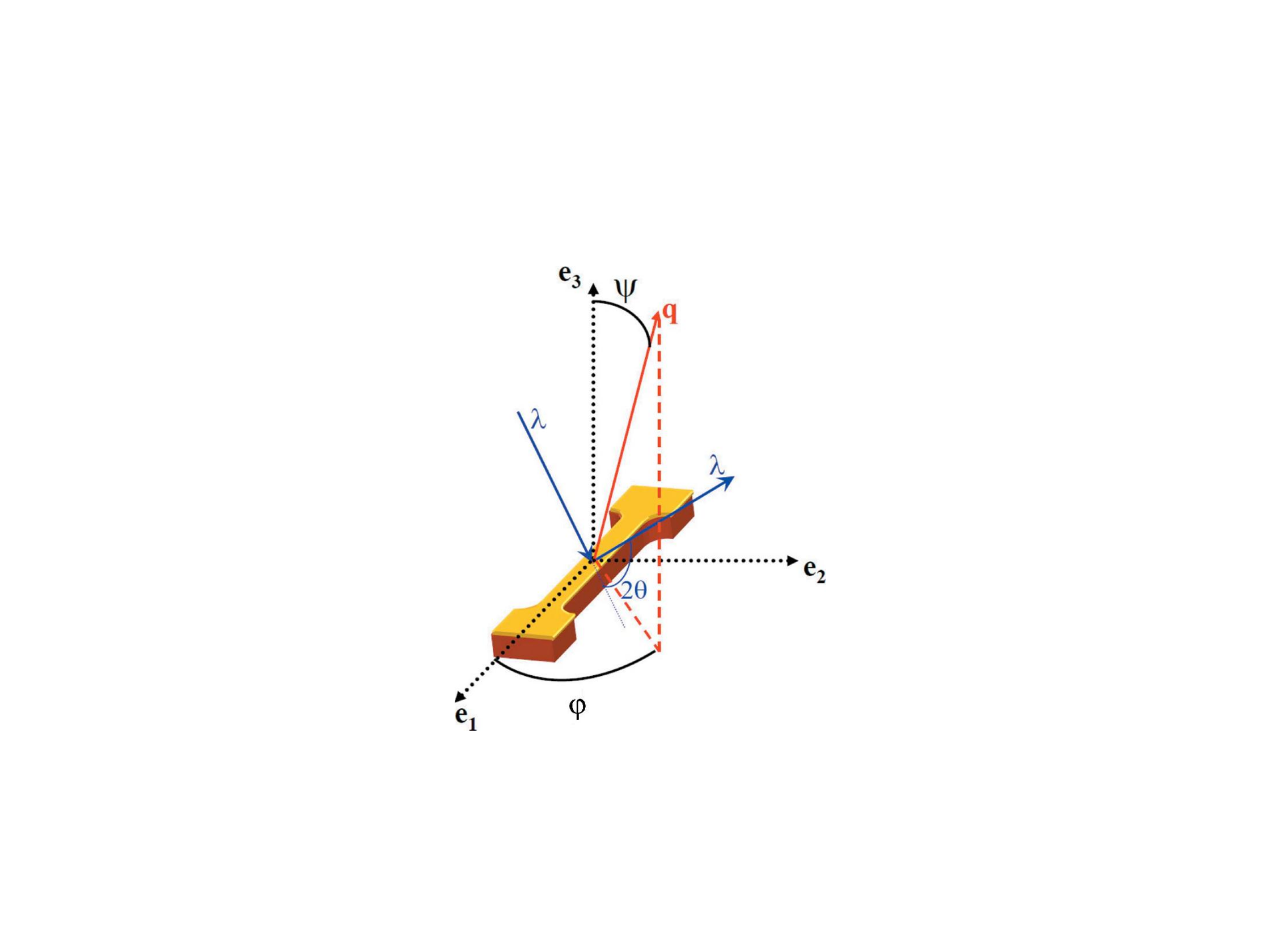}
\protect\caption{\emph{The uniaxial tensile stress applied to the film-substrate system is along $e_{1}$ axis. 
The direction of the scattering vector q is defined by the in-plane azimuthal angle $\phi$ from $e_{1}$ and the
 polar angle $\psi$ from the sample normal direction} \emph{$e_{3}$. $\lambda$ is the wavelength of the x-rays.} \cite{key-7}} 
\end{figure}

A more general linear relation between strain and stress is the generalized
Hooke's law establishing the proportionality between stress and strain
magnitudes, represented in tensor form by \cite{key-12} :

\begin{equation}
\sigma_{ij}=\Lambda_{ijkl}\varepsilon_{kl}
\end{equation}

Here $\sigma_{ij}$, $\Lambda_{ijkl}$ and $\varepsilon_{kl}$  represent the stress tensor, the elasticity tensor and the strain
tensor, respectively. The stress, $\sigma_{ij}$, is a second-order tensor
and it is represented by a square matrix, where the principal diagonal,
$\sigma_{ii}$ are called principal stress and the non-diagonal components
are called shear stress. The strains, $\varepsilon_{kl}$, are
a second-order symmetric tensor . The elasticity tensor, $\Lambda_{ijkl}$,
is a fourth-order tensor and contains $3^{4}$ = 81 elements. . Due
to the symmetry tensor properties we can reduce the number of elastic
constants to only 21. They are called triclinic due to the fact that they have no planes
of symmetry. When the material allows symmetry operations (rotations,
reflections or inversions) in its structure, the number of material
constants is also reduced \cite{key-13} .

 Orthotropic materials have three planes/axes of symmetry.
The elastic behavior of an orthotropic material is characterized by nine independent constants.
In phenomenological terms they are referred to as three longitudinal
moduli of elasticity or Young \textquoteright s moduli ($E_{x}$, $E_{y}$
and $E_{z}$ ), which describe the tendency of an object to deform
along an axis when opposing forces are applied along that axis, three
transverse moduli of elasticity or shear moduli ($G_{xy}$, $G_{yz}$
and $G_{xz}$) which describe an object\textquoteright s tendency
to shear, (the deformation of a shape at a constant volume) when acted
on by opposite forces; and three Poisson coefficients ($\nu_{xy}$,
$\nu_{xz}$ and $\nu_{yz}$) which describe the ratio of transverse
contraction strain to longitudinal extension strain in the direction
of the stretching force \cite{key-15}. Given the measurement of all the engineering constants of an orthotropic material
along the principal directions using the compliance matrix, the elastic
behavior of an orthotropic material is represented as follows, in
Voigt notation \cite{key-13} :

\begin{widetext}

\begin{equation}
\left(\begin{array}{c}
\varepsilon_{xx}\\
\varepsilon_{yy}\\
\varepsilon_{zz}\\
\varepsilon_{yz}\\
\varepsilon_{xz}\\
\varepsilon_{xy}
\end{array}\right)=\left(\begin{array}{cccccc}
\frac{1}{E_{x}} & \frac{-\upsilon_{yx}}{E_{y}} & \frac{-\upsilon_{zx}}{E_{z}} & 0 & 0 & 0\\
\frac{-\upsilon_{xy}}{E_{x}} & \frac{1}{E_{y}} & \frac{-\upsilon_{zy}}{E_{z}} & 0 & 0 & 0\\
\frac{-\upsilon_{xz}}{E_{x}} & \frac{-\upsilon_{yz}}{E_{y}} & \frac{1}{E_{z}} & 0 & 0 & 0\\
0 & 0 & 0 & \frac{1}{2Gyz} & 0 & 0\\
0 & 0 & 0 & 0 & \frac{1}{2G_{xz}} & 0\\
0 & 0 & 0 & 0 & 0 & \frac{1}{2G_{xy}}
\end{array}\right).\left(\begin{array}{c}
\sigma_{xx}\\
\sigma_{yy}\\
\sigma_{zz}\\
\tau_{yz}\\
\tau_{xz}\\
\tau_{xy}
\end{array}\right)
\end{equation}

\end{widetext}

We can obtain the following relations from the principal axis:

\begin{widetext}

\begin{equation}
\left(\begin{array}{c}
\varepsilon_{xx}\\
\varepsilon_{yy}\\
\varepsilon_{zz}
\end{array}\right)=\left(\begin{array}{ccc}
\frac{1}{E_{x}} & \frac{-\upsilon_{yx}}{E_{y}} & \frac{-\upsilon_{zx}}{E_{z}}\\
\frac{-\upsilon_{xy}}{E_{x}} & \frac{1}{E_{y}} & \frac{-\upsilon_{zy}}{E_{z}}\\
\frac{-\upsilon_{xz}}{E_{x}} & \frac{-\upsilon_{yz}}{E_{y}} & \frac{1}{E_{z}}
\end{array}\right).\left(\begin{array}{c}
\sigma_{xx}\\
\sigma_{yy}\\
\sigma_{zz}
\end{array}\right)
\end{equation}
.
\end{widetext}

As the matrix that characterizes the material is symmetric \cite{key-15} , then we have the following relations:

\begin{equation}
\frac{\upsilon_{yx}}{E_{y}}=\frac{\upsilon_{xy}}{E_{x}};\frac{\upsilon_{zx}}{E_{z}}=\frac{\upsilon_{xz}}{E_{x}};\frac{\upsilon_{yz}}{E_{y}}=\frac{\upsilon_{zy}}{E_{z}};
\end{equation}

To calculate the state equation of principal triaxial stress, we use the
relations (4), taking into account the principal directions. Then:

\begin{equation}
\frac{\sigma_{xx}}{E_{x}}-\frac{\upsilon_{yx}}{E_{y}}\sigma_{yy}-\frac{\upsilon_{zx}}{E_{z}}\sigma_{zz}=\varepsilon_{xx}
\end{equation}

\begin{equation}
-\frac{\upsilon_{xy}}{E_{x}}\sigma_{xx}+\frac{\sigma_{yy}}{E_{y}}-\frac{\upsilon_{zy}}{E_{z}}\sigma_{zz}=\varepsilon_{yy}
\end{equation}

\begin{equation}
-\frac{\upsilon_{xz}}{E_{x}}\sigma_{xx}-\frac{\upsilon_{yz}}{E_{y}}\sigma_{yy}+\frac{\sigma_{zz}}{E_{z}}=\varepsilon_{zz}
\end{equation}

In cases where the shear modules are not found in the equations, since they are in the principal axis, we have the following relations \cite{key-11}:

\begin{equation}
G_{xy}=\frac{E_{x}E_{y}}{E_{x}+E_{y}+2\upsilon_{xy}E_{x}}
\end{equation}

\begin{equation}
G_{yz}=\frac{E_{y}E_{z}}{E_{y}+E_{z}+2\upsilon_{yz}E_{y}}
\end{equation}

\begin{equation}
G_{xz}=\frac{E_{x}E_{z}}{E_{x}+E_{z}+2\upsilon_{xz}E_{x}}
\end{equation}

\section{OUR PROPOSAL}

\subsection{ RELATION BETWEEN STRESSES AND  ELASTIC CONSTANTS WITH ORTHOTROPIC SYMMETRY.}

Note that from the equations (6), (7) and (8) we could make algebraic operations
in order to obtain relations between stresses and the elastic
constants. Assuming that the force applied to the thin film is at the
x direction, we can divide the equation (7) by (6) and we obtain:

\begin{equation}
\frac{\sigma_{zz}}{\sigma_{yy}}=\frac{\left(1-\nu_{xy}\nu_{yx}\right)}{\left(\nu_{yz}+\nu_{xz}\nu_{yx}\right)}
\end{equation}

Proceeding analogously, combining equations (8) and (6), and (8) and (7), we get, respectively,

\begin{equation}
\frac{E_{z}}{E_{y}}=\frac{\nu_{xy}\left(1-\nu_{xy}\nu_{yx}\right)}{\left(\nu_{yz}^{2}\nu_{xy}+\nu_{xz}^{2}\nu_{yx}+2\nu_{yz}\nu_{xz}\nu_{xy}\nu_{yx}\right)}
\end{equation}

\begin{equation}
\frac{E_{z}}{E_{x}}=\frac{\nu_{yx}\left(1-\nu_{xy}\nu_{yx}\right)}{\left(\nu_{yz}^{2}\nu_{xy}+\nu_{xz}^{2}\nu_{yx}+2\nu_{yz}\nu_{xz}\nu_{xy}\nu_{yx}\right)}
\end{equation}

In equation (4) we can calculate the determinant of the matrix A associated
with the material. Thus, together with the symmetry relation (5) implies:

\begin{widetext}

\begin{equation}
detA=\frac{\nu_{xy}\left(1-\nu_{xy}\nu_{yx}\right)E_{y}-\left(\nu_{yz}^{2}\nu_{xy}+\nu_{xz}^{2}\nu_{yx}+2\nu_{yz}\nu_{xz}\nu_{xy}\nu_{yx}\right)E_{z}}{\nu_{yx}E_{x}E_{y}E_{x}E_{z}}
\end{equation}

\end{widetext}

Using the relation (13) we obtain that:

\begin{equation}
detA=0
\end{equation}

where A is material matrix. It's indicating that the matrix is not invertible, therefore it can proposal a new equation. 

The equation below is "ad hoc" and obeys the equations (6), (7) and (8).

\begin{equation}
\frac{\sigma_{xx}}{\sigma_{yy}}=\frac{\left(1-\nu_{xy}\nu_{yx}\right)}{\left(\nu_{yz}-\nu_{xz}\nu_{yx}\right)}
\end{equation}

Using equations (12) and (17) in equations (6), (7) and (8), we obtain:

\begin{widetext}

\begin{equation}
\sigma_{xx}=\frac{\left(\nu_{yz}+\nu_{xz}\nu_{yx}\right)\left(1-\nu_{xy}\nu_{yx}\right)\varepsilon_{xx}E_{x}}{\left(\nu_{yz}+\nu_{xz}\nu_{yx}\right)\left(1-\nu_{xy}\nu_{yx}\right)-\left(\nu_{yz}-\nu_{xz}\nu_{yx}\right)\left(\nu_{xz}+\nu_{xy}\nu_{yz}\right)}
\end{equation}

\begin{equation}
\sigma_{yy}=\frac{\left(\nu_{yz}+\nu_{xz}\nu_{yx}\right)\left(\nu_{yz}-\nu_{xz}\nu_{yx}\right)\varepsilon_{yy}E_{y}}{\left(\nu_{yz}+\nu_{xz}\nu_{yx}\right)\left(\nu_{yz}-\nu_{xz}\nu_{yx}\right)-\left(1-\nu_{xy}\nu_{yx}\right)\left[\nu_{yx}\left(\nu_{yz}+\nu_{xz}\nu_{yx}\right)+\nu_{yz}\left(\nu_{yz}-\nu_{xz}\nu_{yx}\right)\right]}
\end{equation}

\begin{equation}
\sigma_{zz}=\frac{\left(1-\nu_{xy}\nu_{yx}\right)\left(\nu_{yz}-\nu_{yx}\nu_{xz}\right)E_{z}\varepsilon_{zz}}{\left(1-\nu_{xy}\nu_{yx}\right)\left(\nu_{yz}-\nu_{xz}\nu_{yx}\right)-\left(\nu_{yz}+\nu_{xz}\nu_{yx}\right)\left[\nu_{zx}\left(1-\nu_{xy}\nu_{yx}\right)+\nu_{zy}\left(\nu_{yz}-\nu_{xz}\nu_{yx}\right)\right]}
\end{equation}

\end{widetext}

The strain and stress tensors are always related to the principal axes.
Substituting relations (6), (7) and (8) in equation (1), we obtain the following result:

\begin{widetext}

\begin{align}
\varepsilon_{\text{\ensuremath{\Phi},\ensuremath{\Psi}}} & =\sin{}^{2}\psi\Bigg\{\frac{\sigma_{xx}}{E_{x}}\left[\left(1+\upsilon_{xy}\right)\cos^{2}\text{\ensuremath{\Phi}}+\left(\upsilon_{xz}-\upsilon_{xy}\right)\right]+\frac{\sigma_{yy}}{E_{y}}\left[\left(1+\upsilon_{yx}\right)\sin^{2}\text{\ensuremath{\Phi}}+\left(\upsilon_{yz}-\upsilon_{yx}\right)\right]\nonumber \\
 & -\frac{\sigma_{zz}}{E_{z}}\left(1+\upsilon_{zy}\sin^{2}\text{\ensuremath{\Phi}}+\upsilon_{zx}\cos^{2}\text{\ensuremath{\Phi}}\right)\Bigg\}-\left(\frac{\upsilon_{xz}\sigma_{xx}}{E_{x}}+\frac{\upsilon_{yz}\sigma_{yy}}{E_{y}}-\frac{\sigma_{zz}}{E_{z}}\right)
\end{align}

\end{widetext}

Where, $\nu_{xy}$, $\nu_{yx}$, $\nu_{xz}$ ,$\nu_{zx}$, $\nu_{yz}$
and $\nu_{zy}$ are the Poisson coefficients, $E_{x}$, $E_{y}$ and
$E_{z}$ are Young\textquoteright s moduli in the x, y, and z
and $\sigma_{xx}$ , $\sigma_{yy}$ and $\sigma_{zz}$ are the principal
stresses at the directions x, y and z, respectively. Equation above
is the key relation used in the model proposed for the principal triaxial
stress state \cite{key-8} .

In what follous we will find a new equation that relates the elastic constants
of the thin film with the force applied to the film-substrate system. 

\subsection{AN EQUATION FOR THE FILM-SUBSTRATE SYSTEM.}

In this section we consider a new equation for the stresses of film-substrate
system which is proposed based on Newton's laws and energy conservation.
We assume the interface between the two-lamina composite (the film and the substrate) which
is subjected to an uniaxial tension load along with the direction $e_{1}$, (Fig 1.) to be perfect \cite{key-7} .
We will assume that, due to the "Poisson's effect" the strain on the z-diection exist and it is not neglectable. Thus, every state of strain and stress are present on the thin film: three principal strains and three principal stresses.

\begin{equation}
\varepsilon_{xx}^{f}=\varepsilon_{xx}^{s}
\end{equation}

\begin{equation}
\varepsilon_{yy}^{f}=\varepsilon_{yy}^{s}
\end{equation}

Using the equation (12) we can relate the $\varepsilon_{zz}$ deformations
in the $\varepsilon_{xx}$ and $\varepsilon_{yy}$ directions, as
follows: 

\begin{equation}
\varepsilon_{zz}^{f}=\frac{\nu_{yz}^{f}}{\nu_{yx}^{f}}\varepsilon_{xx}^{f}-\left(\frac{\sigma_{yy}}{\sigma_{zz}}\right)^{f}\left(\frac{\varepsilon_{xx}^{f}+\nu_{yx}^{f}\varepsilon_{yy}^{f}}{\nu_{yx}^{f}}\right)
\end{equation}

Assuming that the substrate is an orthotropic material, we have:

\begin{equation}
\varepsilon_{zz}^{s}=\frac{\nu_{yz}^{s}}{\nu_{yx}^{s}}\varepsilon_{xx}^{s}-\left(\frac{\sigma_{yy}}{\sigma_{zz}}\right)^{s}\left(\frac{\varepsilon_{xx}^{s}+\nu_{yx}^{s}\varepsilon_{yy}^{s}}{\nu_{yx}^{s}}\right)
\end{equation}
 
We can divide the equation (22) by (23), we get:

\begin{equation}
\nu_{xy}^{f}=\nu_{xy}^{s}
\end{equation}

\begin{equation}
\nu_{yx}^{f}=\nu_{yx}^{s}
\end{equation}

 If the interface between the two laminas is assumed to be perfect,
that is, the resultant load, $F_{app}$, along the longitudinal direction
supported by the speciment film-substrate system is the sum of the loads supported
by the film and the substrate. The forces acting upon the film along the directions x, y and z, we have:

\begin{equation}
F_{app}=F_{x}^{f}+F_{x}^{s}
\end{equation}

\begin{equation}
F_{y}=F_{y}^{f}+F_{y}^{s}=0
\end{equation}

\begin{equation}
F_{z}=F_{z}^{f}+F_{z}^{s}=0
\end{equation}

In terms of stresses:

\begin{equation}
\sigma_{app}=g_{f}\sigma_{xx}^{f}+g_{s}\sigma_{xx}^{s}
\end{equation}

\begin{equation}
\sigma_{yy}=g_{f}\sigma_{yy}^{f}+g_{s}\sigma_{yy}^{s}=0
\end{equation}

\begin{equation}
\sigma_{zz}=g_{f}\sigma_{zz}^{f}+g_{s}\sigma_{zz}^{s}=0
\end{equation}

Where $g_{f}$ and $g_{s}$ are the fractional thicknesses of the
thin film and the substrate, respectively.

With the equations (32) and (33), we get:

\begin{equation}
\left(\frac{\sigma_{yy}}{\sigma_{zz}}\right)^{f}=\left(\frac{\sigma_{yy}}{\sigma_{zz}}\right)^{s}
\end{equation}

Thus the equation (25) yields:

\begin{equation}
\varepsilon_{zz}^{s}=\frac{\nu_{yz}^{f}}{\nu_{yx}^{f}}\varepsilon_{xx}^{f}-\left(\frac{\sigma_{yy}}{\sigma_{zz}}\right)^{f}\left(\frac{\varepsilon_{xx}^{f}+\nu_{yx}^{f}\varepsilon_{yy}^{f}}{\nu_{yx}^{f}}\right) = \varepsilon_{zz}^{f}
\end{equation}

As a result of the above, we can deduce that:

\begin{equation}
\nu_{yz}^{f}=\nu_{yz}^{s}
\end{equation}

The work done by the stresses during the process of deformation is
stored in the form of elastic potential energy \cite{key-15} .

\begin{equation}
dw=\intop\sigma d\varepsilon dv
\end{equation}

For a material with linear behavior, the specific strain energy is
represented as follows:

\begin{equation}
U_{0}=\frac{dw}{dv}=\frac{\sigma\varepsilon}{2}
\end{equation}

In the case of thin films, and considering linear elastic behavior,
the specific strain energy is given by:

\begin{widetext}

\begin{equation}
U_{0}=\frac{1}{2}\sigma_{xx}^{f}\varepsilon_{xx}^{f}+\frac{1}{2}\sigma_{xx}^{s}\varepsilon_{xx}^{s}+\frac{1}{2}\sigma_{yy}^{f}\varepsilon_{yy}^{f}+\frac{1}{2}\sigma_{yy}^{s}\varepsilon_{yy}^{s}+\frac{1}{2}\sigma_{zz}^{f}\varepsilon_{zz}^{f}+\frac{1}{2}\sigma_{zz}^{s}\varepsilon_{zz}^{s}
\end{equation}

\end{widetext}

Where f end s indices are related to the film and the substrate, respectively.

Using equations (22), (23) and (35) and then (32),(33) and (34)
in the equation above, we have:

\begin{widetext}

\begin{equation}
U_{0}=\frac{1}{2g_{s}}\sigma_{xx}^{f}\left(g_{s}-g_{f}\right)\varepsilon_{xx}^{f}+\frac{1}{2g_{s}}\sigma_{app}\varepsilon_{xx}^{f}+\frac{1}{2g_{s}\nu_{yx}^{f}}\sigma_{yy}^{f}\left(g_{f}-g_{s}\right)\varepsilon_{xx}^{f}+\frac{1}{2g_{s}\nu_{yx}^{f}}\sigma_{zz}^{f}\nu_{yz}^{f}\left(g_{s}-g_{f}\right)\varepsilon_{xx}^{f}
\end{equation}

\end{widetext}

This equation relates the specific strain energy for the thin film
with respect to stress and strain in the film. Let us suppress the index
f since all terms are related to the film.

According to Drucker's postulate \cite{key-14} , the work done by the external agent
during the deformation process is stored in the form of elastic potential energy, meaning that.

\begin{equation}
W_{ext}=W_{all}
\end{equation}

\begin{equation}
U_{0}=\frac{W_{all}}{V}=2\sigma_{app}.\varepsilon_{all}
\end{equation}

The total strain on the film-substrate system is: 

\begin{equation}
\varepsilon_{all}=\varepsilon_{all}^{f}+\varepsilon_{all}^{s}
\end{equation}

Where $\varepsilon_{all}$ is the sum of the total deformation module
of the film and substrate and depends on the geometry.

Manipulating (40), (42) and (43) algebracaly we obtained:

\begin{widetext}

\begin{equation}
\sigma_{app}=\left\{ \frac{\left(g_{f}-g_{s}\right)\left(-\nu_{yx}\sigma_{xx}+\sigma_{yy}-\nu_{yz}\sigma_{zz}\right)}{\nu_{yx}\left[4g_{s}\left(\varepsilon_{all}^{f}+\varepsilon_{all}^{s}\right)-\varepsilon_{xx}\right]}\right\} \varepsilon_{xx}^{f}
\end{equation}

\end{widetext}

This equation expresses the stresses to the thin film due to the
stresses applied on the film-substrate system for orthotropic materials.

Using equation (7) and the properties of symmetry (5) we obtain:

\begin{widetext}

\begin{equation}
E_{y}=\left\{ \frac{\left[4g_{s}\left(\varepsilon_{all}^{f}+\varepsilon_{all}^{s}\right)-\varepsilon_{xx}\right]}{\left(g_{f}-g_{s}\right)\varepsilon_{xx}}\right\} \frac{\sigma_{app}}{\varepsilon_{yy}}\nu_{yx}
\end{equation}

\end{widetext}

This equation relates the Young's modulus of the film in the y direction
to the stresses  applied to the film-substrate system for orthotropic material . Let us now make an
application of the obtained equations.

\subsection{SET OF EQUATIONS WITH TRANSVERSE ISOTROPIC SYMMETRY.}

Note that a material with transversely isotropic anisotropy has
five constants that characterize its elastic tensor. Assuming an xy
plane of symmetry, the elastic constants are $E_{x}$,
$E_{z}$, $G_{xy}$, $\nu_{xy}$ and $\nu_{xz}$.

Equations (6), (7) and (8), for transversely isotropic symmetry, $( \nu_{xy}=  \nu_{yx})$,  become:

\begin{equation}
\frac{\sigma_{xx}}{E_{x}}-\frac{\upsilon_{xy}}{E_{x}}\sigma_{yy}-\frac{\upsilon_{xz}}{E_{x}}\sigma_{zz}=\varepsilon_{xx}
\end{equation}

\begin{equation}
-\frac{\upsilon_{xy}}{E_{x}}\sigma_{xx}+\frac{\sigma_{yy}}{E_{x}}-\frac{\upsilon_{xz}}{E_{x}}\sigma_{zz}=\varepsilon_{yy}
\end{equation}

\begin{equation}
-\frac{\upsilon_{xz}}{E_{x}}\sigma_{xx}-\frac{\upsilon_{xz}}{E_{x}}\sigma_{yy}+\frac{1}{E_{z}}\sigma_{zz}=\varepsilon_{zz}
\end{equation}

With these results we can calculate the coefficients of the Poisson,
assuming that the force applied to the thin film is the x direction. 

\begin{equation}
\upsilon_{xy}=-\frac{\varepsilon_{yy}}{\varepsilon_{xx}};\upsilon_{xz}=-\frac{\varepsilon_{zz}}{\varepsilon_{xx}}
\end{equation}

Using the assumption that the material is transversely isotropic,
equations (12), (13), (17), (18), (19) and (20) then reduce to the following ones:

\begin{equation}
\frac{\sigma_{zz}}{\sigma_{yy}}=\frac{\left(1-\upsilon_{xy}\right)}{\upsilon_{xz}}
\end{equation}

\begin{equation}
\frac{\sigma_{xx}}{\sigma_{yy}}=\frac{\left(1+\upsilon_{xy}\right)}{\upsilon_{xz}}
\end{equation}

\begin{equation}
\frac{E_{z}}{E_{x}}=\frac{\left(1-\upsilon_{xy}\right)}{2\upsilon_{xz}^{2}}
\end{equation}

\begin{equation}
\sigma_{xx}=\frac{\left(1+\upsilon_{xy}\right)}{\left(1+\upsilon_{xy}-\upsilon_{xz}\right)}\varepsilon_{xx}E_{x}
\end{equation}

\begin{equation}
\sigma_{yy}=-\frac{\upsilon_{xz}}{\upsilon_{xy}\left(1+\upsilon_{xy}-\upsilon_{xz}\right)}\varepsilon_{yy}E_{x}
\end{equation}

\begin{equation}
\sigma_{zz}=-\frac{\left(1-\upsilon_{xy}\right)}{\upsilon_{xz}\left(1+\upsilon_{xy}-\upsilon_{xz}\right)}\varepsilon_{zz}E_{x}
\end{equation}

Subtracting the relations  (53) and (54), we have:

\begin{equation}
(\sigma_{xx}-\sigma_{yy})=\frac{E_{x}}{(1+\upsilon_{xy})}(\varepsilon_{xx}-\varepsilon_{yy})
\end{equation}

Using the maximum shear theory condition  \cite{key-16}: 

\begin{equation}
(\sigma_{xx}-\sigma_{yy})=2G_{xy}(\varepsilon_{xx}-\varepsilon_{yy})
\end{equation}

Comparing the equations (56) and (57), we obtained:

\begin{equation}
G_{xy}=\frac{E_{x}}{2(1+\upsilon_{xy})}
\end{equation}

Where $G_{xy}$ is the shear modulus in the xy plane.

For the conditions transversely isotropic the equation (21) becomes:

\begin{widetext}

\begin{align}
\varepsilon_{\text{\ensuremath{\Phi},\ensuremath{\Psi}}} & =\sin{}^{2}\psi\Bigg\{\frac{\sigma_{xx}}{E_{x}}\left[\left(1+\upsilon_{xy}\right)\cos^{2}\text{\ensuremath{\Phi}}+\left(\upsilon_{xz}-\upsilon_{xy}\right)\right]+\frac{\sigma_{yy}}{E_{x}}\left[\left(1+\upsilon_{xy}\right)\sin^{2}\text{\ensuremath{\Phi}}+\left(\upsilon_{xz}-\upsilon_{xy}\right)\right]\nonumber \\
 & -\frac{\sigma_{zz}}{E_{z}}\left(1+\upsilon_{zx}\sin^{2}\text{\ensuremath{\Phi}}+\upsilon_{zx}\cos^{2}\text{\ensuremath{\Phi}}\right)\Bigg\}-\left(\frac{\upsilon_{xz}\sigma_{xx}}{E_{x}}+\frac{\upsilon_{xz}\sigma_{yy}}{E_{x}}-\frac{\sigma_{zz}}{E_{z}}\right)
\end{align}

\end{widetext}

\subsection{AN EQUATION FOR THE FILM-SUBSTRATE SYSTEM WITH TRANSVERSE ISOTROPIC SYMMETRY.}

We assume a two-lamina composite (the film and the substrate) with
transverse isotropic symmetry which is subjected to a uniaxial tension
load along the direction $s_{1}$. If the interface between the two
laminas is assumed to be perfect, that is, the resultant load $F_{app}$
along the longitudinal direction suported by the speciment film-substrate system
is the sum of the loads supported by the film and the substrate:

Using the equation (24), for orthotropic materials, we can relate
the $\varepsilon_{zz}$ deformations in the $\varepsilon_{xx}$ and
$\varepsilon_{yy}$ directions, as follows: 

\begin{equation}
\varepsilon_{zz}^{f}=-\left(\frac{\sigma_{yy}}{\sigma_{zz}}\right)^{f}\left(\varepsilon_{xx}^{f}+\varepsilon_{yy}^{f}\right)
\end{equation}

Assuming that the substrate is a transversely isotropic materials:

\begin{equation}
\varepsilon_{zz}^{s}=-\left(\frac{\sigma_{yy}}{\sigma_{zz}}\right)^{s}\left(\varepsilon_{xx}^{s}+\varepsilon_{yy}^{s}\right)
\end{equation}

and equation (44) becomes:

\begin{widetext}

\begin{equation}
\left(\sigma_{xx}-\sigma_{yy}\right)=\left\{ \frac{\left[12g_{s}\left(\sqrt{1+\left(\frac{\varepsilon_{yy}}{\varepsilon_{xx}}\right)^{2}+\left(\frac{\varepsilon_{zz}}{\varepsilon_{xx}}\right)^{2}}\right)-1\right]}{\left(g_{s}-g_{f}\right)}\right\} \sigma_{app}
\end{equation}

\end{widetext}

Now, subtracting the equations (53) and (54) and comparing to the
above equation, we get:

\begin{equation}
\left(\sigma_{xx}-\sigma_{yy}\right)=\varepsilon_{xx}E
\end{equation}

Altogether, this implies that:

\begin{widetext}

\begin{equation}
E=\left\{ \frac{\left[12g_{s}\left(\sqrt{1+\left(\frac{\varepsilon_{yy}}{\varepsilon_{xx}}\right)^{2}+\left(\frac{\varepsilon_{zz}}{\varepsilon_{xx}}\right)^{2}}\right)-1\right]}{\left(g_{s}-g_{f}\right)}\right\} \frac{\sigma_{app}}{\varepsilon_{xx}}
\end{equation}

\end{widetext}

This equation relates the Young's modulus of a transversely  isotropic film with the stresses applied to the film-substrate system.

\section{DATA APPLICATION ON TRANSVERSELY ISOTROPIC EQUATIONS.}

\subsection{ FIBER-TEXTURED THIN GOLD FILM.}

We consider the data of Faurie et al. (2005), which establishes the
elastic constants of a fiber-textured thin gold film with transversely
isotropic symmetry by combining synchrotron X-ray diffraction and
in situ tensile testing.Gold thin films save been deposited by physical-vapor deposition. The substrate was a 127.5-$\mu$m-thick polymide dog-bone substrate; the in-plane sample dimensions were 14 X 6 $mm^2$.
 A  surface profilometer system to be 700 $\pm$ 10 nm \cite{key-7}.

Once the experimental values are plotted ($\Psi=0$\textdegree{}
, $\Phi$ = 0\textdegree{} ) and ($\Psi=0$\textdegree{} , $\Phi$ =
90\textdegree ), as shown in figure 2, we obtain the following empirical
equations:

\begin{equation}
\varepsilon_{0,0}=-1,1.10^{-4}F
\end{equation}

\begin{equation}
\varepsilon_{90,0}=-1,1.10^{-4}F
\end{equation}

Where F is the applied force.

Then, using equation (59), we obtain the following theoretical equation: 

\begin{equation}
\varepsilon_{0,0}=-\left[\frac{\upsilon_{xz}}{E}+\left(\frac{\sigma_{yy}}{\sigma_{xx}}\right)\frac{\upsilon_{xz}}{E}-\left(\frac{\sigma_{zz}}{\sigma_{xx}}\right)\frac{1}{E_{z}}\right]\frac{F}{B}
\end{equation}

\begin{equation}
\varepsilon_{90,0}=-\left[\frac{\upsilon_{xz}}{E}+\left(\frac{\sigma_{yy}}{\sigma_{xx}}\right)\frac{\upsilon_{xz}}{E}-\left(\frac{\sigma_{zz}}{\sigma_{xx}}\right)\frac{1}{E_{z}}\right]\frac{F}{B}
\end{equation}

Where $F=B\sigma_{xx}$ and B is constant.

\begin{figure}[!h]
\includegraphics[scale=0.5]{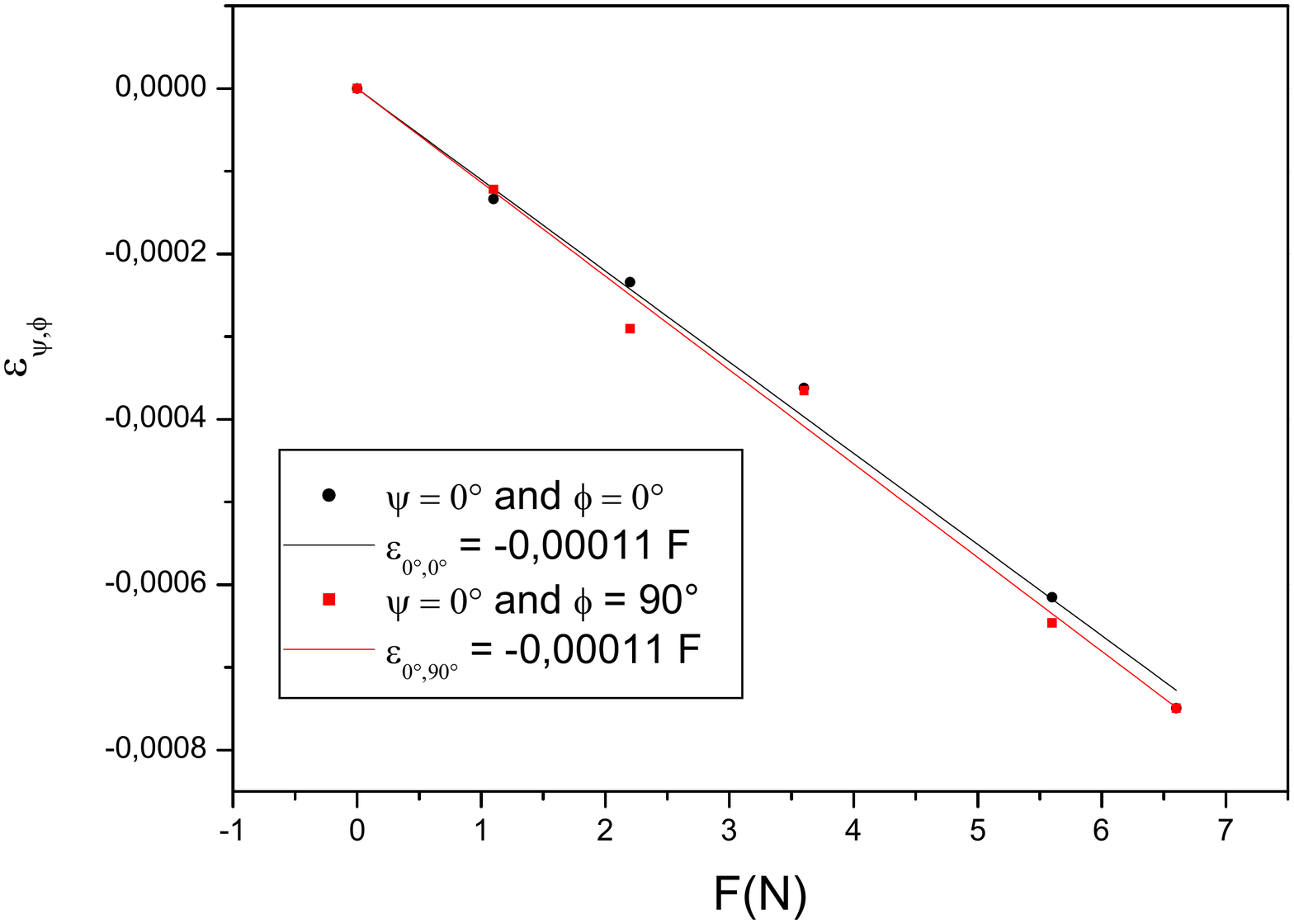}

\protect\caption{\emph{Graph of strain versus force applied to the film for $\varPsi$=0\textdegree .
The black dots represent $\phi$= 0\textdegree{} and the red dots
represent $\phi$= 90\textdegree .}}
\end{figure}

Doing the same procedure, now the values$(\Psi=75,04\text{\textdegree},$$\Phi$
= 0\textdegree{} ), $(\Psi=75,04\text{\textdegree},$$\Phi$ = 90\textdegree{}
) and $(\Psi=54,74\text{\textdegree},$$\Phi$ = 0\textdegree{} ), $(\Psi=54,74\text{\textdegree},$$\Phi$
= 90\textdegree{} ), which are shown in Figure 3 and 4 respectively,
we obtain the equations:

\begin{equation}
\begin{aligned}\varepsilon_{\text{0\textdegree\ ; 75,04\textdegree}}=2,2.10^{-4}F\end{aligned}
\end{equation}

\begin{equation}
\varepsilon_{\text{90\textdegree\ ; 75,04\textdegree}}=-1,2.10^{-4}F
\end{equation}

\begin{equation}
\begin{split}\varepsilon_{\text{0\textdegree\ ; 54,74\textdegree}}=1,6.10^{-4}F\end{split}
\end{equation}
\begin{equation}
\varepsilon_{\text{90\textdegree,54,74\textdegree}}=-1,5.10^{-4}F
\end{equation}

\begin{figure}
\includegraphics[scale=0.5]{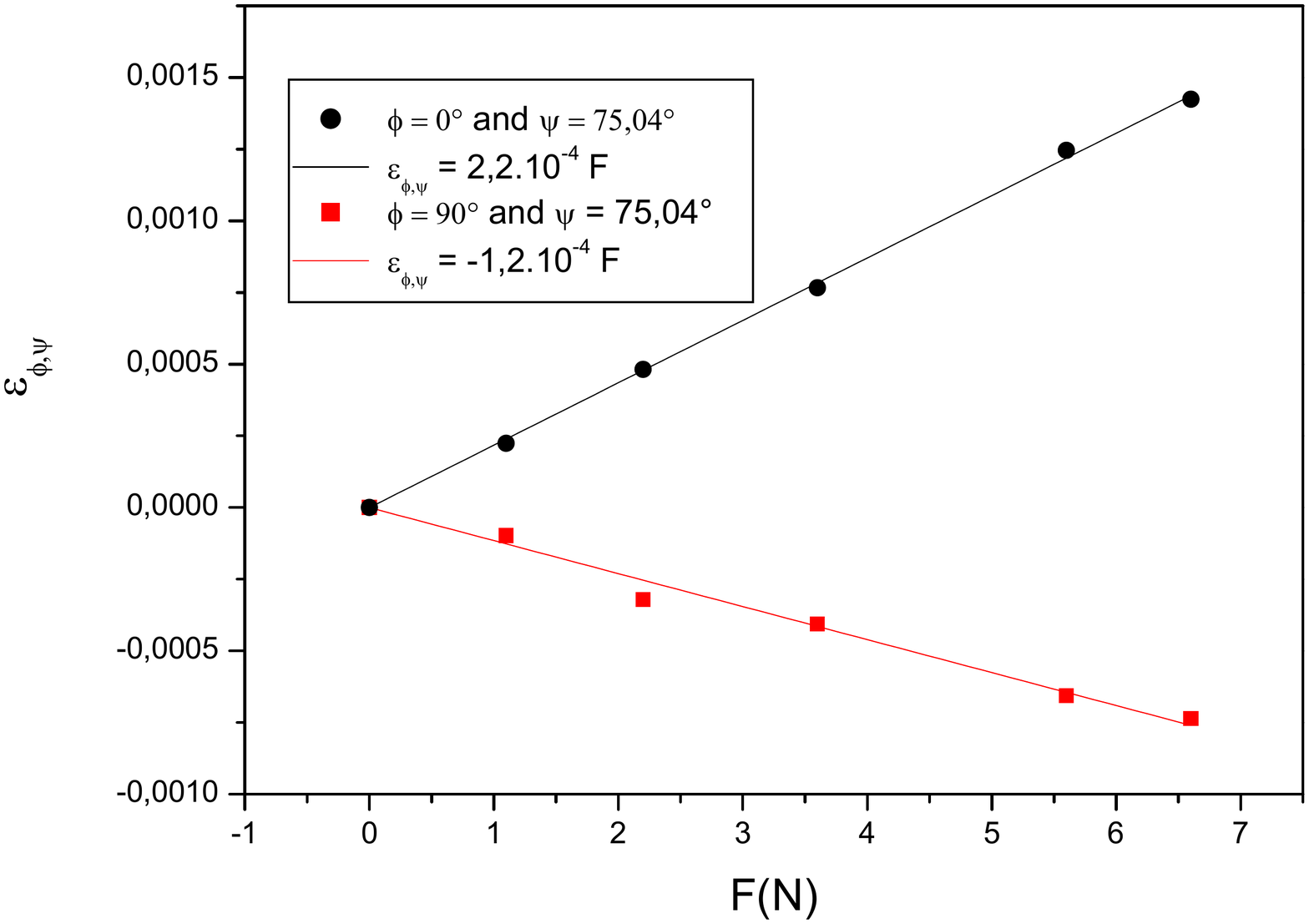}

\protect\caption{\emph{Graph of strain versus force applied to the film for $\varPsi$=75,04\textdegree .
The black dots represent $\phi$= 0\textdegree{} and the red dots
represent $\phi$= 90\textdegree .}}
\end{figure}

\begin{figure}
\includegraphics[scale=0.5]{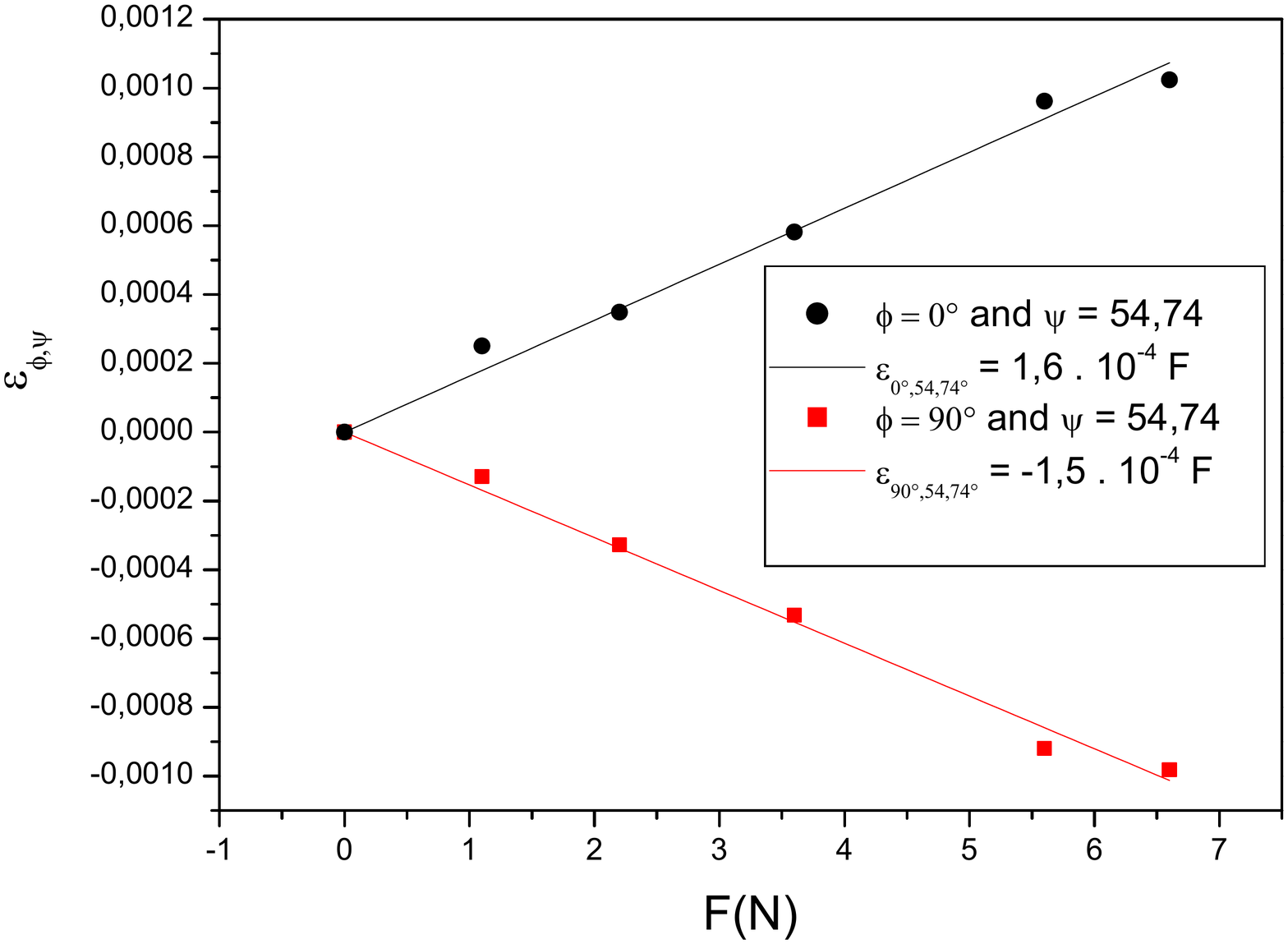}
\protect\caption{\emph{Graph of strain versus force applied to the film for $\varPsi$=54,74\textdegree .
The black dots represent $\phi$= 0\textdegree{} and the red dots
represent $\phi$= 90\textdegree .}}
\end{figure}

We obtain from the equation (59) the following theoretical equations:

\begin{widetext}

\begin{equation}
\varepsilon_{\text{0\textdegree;\text{\ensuremath{\Psi}} }}=\left\{ \left[\frac{\left(1+\upsilon_{xz}\right)}{E}+\left(\frac{\sigma_{yy}}{\sigma_{xx}}\right)\frac{\left(\upsilon_{xz}-\upsilon_{xy}\right)}{E}-\left(\frac{\sigma_{zz}}{\sigma_{xx}}\right)\frac{\left(1+\upsilon_{zx}\right)}{E_{z}}\right]sen^{2}\psi-\left[\frac{\upsilon_{xz}}{E}\left(1+\frac{\sigma_{yy}}{\sigma_{xx}}\right)-\frac{1}{E_{z}}\left(\frac{\sigma_{zz}}{\sigma_{xx}}\right)\right]\right\} \frac{F}{B}
\end{equation}

\begin{equation}
\varepsilon_{\text{90\textdegree;\ensuremath{\Psi}}}=\left\{ \left[\frac{\left(\upsilon_{xz}-\upsilon_{xy}\right)}{E}+\left(\frac{\sigma_{yy}}{\sigma_{xx}}\right)\frac{\left(1+\upsilon_{xz}\right)}{E}-\left(\frac{\sigma_{zz}}{\sigma_{xx}}\right)\frac{\left(1+\upsilon_{zx}\right)}{E_{z}}\right]sen^{2}\psi-\left[\frac{\upsilon_{xz}}{E}\left(1+\frac{\sigma_{yy}}{\sigma_{xx}}\right)+\frac{\upsilon_{xz}\sigma_{yy}}{E}-\frac{1}{E_{z}}\left(\frac{\sigma_{zz}}{\sigma_{xx}}\right)\right]\right\} \frac{F}{B}
\end{equation}
\end{widetext}

Based on the obtained relations, we got the following results, for
F= 6,6 N, shown in the table I:

\begin{table*} [h!]
\begin{tabular}{|c|c|c|c|}
\hline 
$\varepsilon_{xx}(.10^{-4})$($\phi=0\text{\textdegree)}$ & $\varepsilon_{yy}(.10^{-4})$($\phi=90\text{\textdegree)}$ & $\varepsilon_{zz}(.10^{-4})$$(\phi=0\text{\textdegree\ and 90\textdegree})$ & $\Psi$\tabularnewline
\hline 
\hline 
19,21 & -11,84 & - & 54,74\textdegree{}\tabularnewline
\hline 
16,02 & -7,47 & - & 75,04\textdegree{}\tabularnewline
\hline 
- & - & -7,39 & 0\textdegree{}\tabularnewline
\hline 
17,62 & -9,66 & -7,39 & Average \tabularnewline
\hline 
\end{tabular}

\protect\caption{\emph{Result of the average values of the deformations for $\Psi$ angles,
$\Psi=0\text{\textdegree, }\Psi=54,74\text{\textdegree\ and \ensuremath{\Psi}=75,04\text{\textdegree, aplying on film, F = 6,6 N. }}$}}
\end{table*}

With this procedure we obtained $\varepsilon_{xx}$, $\varepsilon_{yy}$
and $\varepsilon_{zz}$ for all force and calculate $\nu_{xy}$ and
$\nu_{xz}$ using equation (49) shown in table II.
\begin{table*} [h!]
\begin{tabular}{|c|c|c|c|}
\hline 
$F(N)$& $\varepsilon_{xx}(.10^{-4})$ & $\varepsilon_{yy}(.10^{-4})$ & $\varepsilon_{zz}(.10^{-4})$\tabularnewline 
\hline 
\hline 
6,6 & 17,62 & -9,66 & -7,39\tabularnewline
\hline 
5,6 & 14,94 & -8,21 & -6,27\tabularnewline
\hline 
 3,6 & 9,61 & -5,27 & -4,03\tabularnewline
\hline 
 2,2 & 5,87 & -3,22 & -2,46\tabularnewline
\hline 
1,1 &  2,94  & -1,61 & -1,23\tabularnewline
\hline 
\end{tabular}

\protect\caption{\emph{The table lists the strain in the film for each applied force}}
\end{table*}

Based on the obtained relations,(50-55),  we have the following results shown in Table III and IV. 
\begin{table*} [h!]
\begin{tabular}{|c|c|c|c|c|c|}
\hline 
$\sigma_{xx}(2005)$(MPa) & $\sigma_{yy}(2005)$(MPa) & F(N) & $\sigma_{xx}(MPa)$ & $\sigma_{yy}(MPa)$ & $\sigma_{zz}(MPa)$\tabularnewline
\hline 
\hline 
135 & 37 & 6,6 & 156 & 42 & 45\tabularnewline
\hline 
115 & 31,4 & 5,6 & 132 & 35,8 & 38,4\tabularnewline
\hline 
73,7 & 20,2 & 3,6 & 85 & 23 & 24,7\tabularnewline
\hline 
45 & 12,3 & 2,2 & 52 & 14 & 15\tabularnewline
\hline 
22,5 & 6 & 1,1 & 26 & 7 & 7,5\tabularnewline
\hline 
\end{tabular}

\protect\caption{\emph{Table to compare the stresses obtained from the equations proposed
and stresses obtained by Faurie (2005).}}
\end{table*}

The other table:
\begin{table*} [h!]
\begin{tabular}{|c|c|c|c|c|c|c|c|}
\hline 
 & E(GPa) & $E_{x}$(GPa) & $E_{z}$(GPa) & $G_{xy}$(GPa) & $\nu$  & $\nu_{xy}$  & $\nu_{xz}$ \tabularnewline
\hline 
\hline 
Faurie (2005) & 75,7 & - & - & - & 0,52 & - & -\tabularnewline
\hline 
Today & - & 64,6 & 82,7 & 20,9 & - & 0,55 & 0,42\tabularnewline
\hline 
\end{tabular}

\protect\caption{\emph{Table to compare the elastic constants obtained from the proposed
equations and the elastic constants obtained by Faurie (2005).}}
\end{table*} 

\section{RESULTS AND DISCUSSION.}

In calculating the determinant of the matrix that characterizes the material, the result is zero, indicating that the matrix is not
invertible. In this calculation the symmetry properties of the matrix of the material, equation (5), were considered .

In Table II, the values of the strains on the z-direction are not neglectables.
The thin film was analyzed in three dimensions and once we take into account the result of the
equation (21), we realize how important the sigma $\sigma_{zz}$ is for its deformation.
The tensions obtained are all positive, the sign of $\sigma_{zz}$ on (21) equation is the inverse of $\sigma_{xx}$ and $\sigma_{yy}$
directions. We note that, in this case, the stress is the contraction.

Equation (17) was obtained "ad hoc", obeys the equations (6), (7) and (8), and the comparison of the results obtained from the stresses in the film with the results obtained by Faurie (2005),
 as seen in table III, is consistent with the values for each of the applied forces.

Looking at equations (44) and (45) we see that these equation
relates the applied stress and the dimensions of the film-substrate system
in the stress on the film. Being possible to calculate the Young's modulus of film.

Looking the deduction of equation (58). The equation (58) is in agreement with the literature.

In table IV, we obtain the five elastic constants, thus completely
characterizing the transversely isotropic material. 

\section{CONCLUSIONS.}

In this work based on elasticity theory of continuous media for small
deformations, it was possible to obtain a set of equations which are able to
completely characterize a thin film on orthotropic symmetry in
their mechanical properties. It was also possible to determine the principal
stresses and strains applied to the thin film. When this set of equations
was applied to a film-substrate system considered transversely isotropic,
the results were consistent with respect to experimental studies of Faurie et al (2005) in what concerns elastic constants. 

\begin{acknowledgments}

The authors gratefully acknowledge the LSPM-CNRS, Universit\'{e} Paris XIII. We would like to thank the
Coordination of Improvement of Higher Education Personnel (CAPES)
foundation grant BEX 9210-13-0.\end{acknowledgments}

\end{document}